*Article*

# TPSQLi: Test Prioritization for SQL Injection Vulnerability Detection in Web Applications

Guan-Yan Yang [1], Farn Wang [1,*], You-Zong Gu [1,2], Ya-Wen Teng [1], Kuo-Hui Yeh [3,4,*], Ping-Hsueh Ho [1] and Wei-Ling Wen [1]

[1] Department of Electrical Engineering, National Taiwan University, Taipei City 106319, Taiwan; f11921091@ntu.edu.tw (G.-Y.Y.); r09921a19@ntu.edu.tw (Y.-Z.G.); r12921059@ntu.edu.tw (Y.-W.T.); r11921a16@ntu.edu.tw (P.-H.H.); r11921100@ntu.edu.tw (W.-L.W.)
[2] CyberLink Corporation, New Taipei City 231023, Taiwan
[3] Institute of Artificial Intelligence Innovation, National Yang Ming Chiao Tung University, Hsinchu City 300093, Taiwan
[4] Department of Information Management, National Dong Hwa University, Hualien 974301, Taiwan
* Correspondence: farn@ntu.edu.tw (F.W.); khyeh@nycu.edu.tw (K.-H.Y.)

**Abstract:** The rapid proliferation of network applications has led to a significant increase in network attacks. According to the OWASP Top 10 Projects report released in 2021, injection attacks rank among the top three vulnerabilities in software projects. This growing threat landscape has increased the complexity and workload of software testing, necessitating advanced tools to support agile development cycles. This paper introduces a novel test prioritization method for SQL injection vulnerabilities to enhance testing efficiency. By leveraging previous test outcomes, our method adjusts defense strength vectors for subsequent tests, optimizing the testing workflow and tailoring defense mechanisms to specific software needs. This approach aims to improve the effectiveness and efficiency of vulnerability detection and mitigation through a flexible framework that incorporates dynamic adjustments and considers the temporal aspects of vulnerability exposure.

## 1. Introduction

This section provides an overview of the paper's background, motivation, contributions, and organization.

### 1.1. Background

Over the past two decades, the integration of network technology into daily life has significantly increased, marking a notable trend [1]. With the growing dependence on the web, web applications have become indispensable for modern data management and processing. This evolution allows developers to efficiently design, maintain, and secure applications by modifying existing code rather than building entire programs from scratch. This streamlined approach not only enhances the protection of valuable business assets, but also transforms the Internet into a vast repository of information. Despite these advancements, web applications remain susceptible to numerous security threats, including Structured Query Language (SQL) injection, cross-site scripting (XSS), and cross-site request forgery (CSRF) [2,3].

Exploiting these vulnerabilities enables attackers to compromise user information and damage organizational reputations by breaching security policies. The primary objective of cybersecurity is to protect networks and resources from unauthorized access, ensuring the principles of integrity, availability, and confidentiality are upheld.

Recent reports, such as the OWASP Top 10 project published in 2021 [4], indicate that injection attacks rank among the top three common vulnerabilities in software projects. SQL injection attacks, in particular, exploit inadequate input validation and poor website management, allowing attackers to inject malicious SQL statements into queries generated by web applications. This can lead to unauthorized access to backend databases and exposure to sensitive information, such as usernames, passwords, email addresses, phone numbers, and credit card details. Additionally, attackers can alter database schemas and content, exacerbating the potential damage.

Various types of SQL injection attacks have been identified [5], and numerous methods for detecting and preventing them have been proposed. However, existing solutions often have limitations and may not fully address the evolving nature of these attacks. As new attack vectors emerge, it is crucial to identify and understand current technologies to develop more effective countermeasures.

*1.2. Motivation*

Software testing is a crucial and costly industry activity, often accounting for over 50% of the total software development cost [6]. To address this challenge in testing SQL injection for agile software development, we propose a software testing tool for automatic testing, activated upon the tester's submission of the software version and relevant information. This tool customizes defense functions tailored to the specific software under test, enhancing the testing process's efficiency. Following the completion of testing, the tool adjusts the defense strength vector to optimize subsequent tests.

Upon receiving the version information, the tool tests the failed cases in the previous round and further designs a choice based on common tester practices. Prioritizing previously failed test cases allows testers to obtain real-time information about iterative versions promptly. After testing, adjustments to the test process are made to facilitate future tests, thus improving overall testing efficiency.

To demonstrate the impact of test prioritization on time and cost, we conducted a pre-test to measure the time differences using various prioritization methods to identify the first SQL injection vulnerabilities. For instance, in target1 (powerhr), the time difference between the TEUQSB and BUSQET prioritization orders was up to 64 s, as shown in Figure 1. Similar differences were observed for target2 (dcstcs) and target3 (healthy), with time differences exceeding 15 s, as illustrated in Figure 1. This substantial variance underscores the importance of optimizing test prioritization strategies. Each letter in the sequences BUSQET represents a different type of SQL injection, detailed in Table 1, with further explanations provided in Section 2.2 SQL Injection.

| powerhr | | dctcs | | healthy | |
|---|---|---|---|---|---|
| TEUQSB | :: 2.0630388259887695 | BQSTEU | :: 4.399706363677978 | TEUSQB | :: 1.9185783863067627 |
| TESUQB | :: 2.07590651512146 | BTQUSE | :: 4.445225715637207 | TEUSBQ | :: 1.9838907718658447 |
| TEUSBQ | :: 2.104346513748169 | BTESQU | :: 4.522237777709961 | TEUBSQ | :: 2.0504517555236816 |
| TEUBSQ | :: 2.1099705696105957 | BUQTES | :: 4.553324460983276 | BEUSQT | :: 2.0795555114746094 |
| TEUQBS | :: 2.1108734607696533 | BQUTSE | :: 4.554885149002075 | UQEBST | :: 2.1025936603546143 |
| BUQSTE | :: 118.60430979728699 | USTEQB | :: 64.93485140800476 | QSTBUE | :: 20.67803692817688 |
| BUSEQT | :: 126.49982142448425 | EUQSTB | :: 65.133220911026 | QSTBEU | :: 22.581323623657227 |
| BUQSET | :: 126.87057614326477 | EUTQSB | :: 65.7765417098999 | QSBETU | :: 24.575214385986328 |
| BUSETQ | :: 131.27171540260315 | UETQSB | :: 66.23421788215637 | QSBEUT | :: 29.8536639213562 |
| BUSQET | :: 132.98175716400146 | EUSQTB | :: 66.56440830230713 | QSBTEU | :: 40.49846863746643 |

**Figure 1.** Pre-test: Different test prioritization leads to different time costs.

**Table 1.** Abbreviation and complete name mapping.

| Complete Name | Abbreviation |
|---|---|
| Boolean-based blind SQL injection | B |
| Union-based SQL injection | U |
| Stack-based SQL injection | S |
| Inline Queries SQL injection | Q |
| Error-based SQL injection | E |
| Time-based blind SQL injection | T |

Additionally, we evaluated the relevant information available in some open-source tools. We found that in ZAP [7], this problem was listed in the TODO section, clearly indicating in the ZAP documentation that this area requires expansion, as shown in Figure 2. Conversely, SQLMAP [8] uses a fixed BEUSTQ order for testing, indicating that SQLMAP does not significantly incorporate test prioritization, as shown in Figure 3. This evidence highlights the need for tools to adopt more dynamic and optimized test prioritization strategies to enhance testing efficiency and effectiveness.

```
339        // Maybe could be a good idea to sort tests
340        // according to the behavior and the heaviness
341        // TODO: define a compare logic and sort it according to that
```

**Figure 2.** Evidence in ZAP.

```
Techniques:
  These options can be used to tweak testing of specific SQL injection
  techniques

  --technique=TECH..  SQL injection techniques to use (default "BEUSTQ")
```

**Figure 3.** Evidence in SQLMAP.

*1.3. Contribution*

In this subsection, we outline the critical contributions of our work:

- We proposed a new algorithm to prioritize SQL Injection vulnerability test cases. This algorithm is part of a comprehensive framework (TPSQLi) that includes the design of weight functions, dynamic adjustments, and evaluation methods to ensure efficient and effective prioritization.
- We enhanced the testing process in SQLMAP by prioritizing test cases that failed in previous rounds, thereby improving the efficiency and effectiveness of the testing cycle.
- Our framework (TPSQLi) is designed to adapt to immediate feedback and evolving security threats, ensuring continuous adjustments to testing priorities. This adaptability significantly enhances the efficiency of security testing, particularly for regression testing, ensuring that testing remains relevant and effective in addressing the ever-changing landscape of web application security.
- Our framework (TPSQLi) performs better than ART4SQLi [9], the current state-of-the-art (SOTA) test prioritization framework, by designing more efficient and adaptable prioritization mechanisms.

*1.4. Organization*

The rest of this paper is organized as follows: Section 2 introduces the preliminary knowledge essential for our study, providing the foundational background. Section 3 offers a comprehensive review of related research, establishing the context and significance of our research. Section 4 presents the TPSQLi penetration testing framework and the test prioritization algorithm unit, detailing the conceptual design and implementation processes with insights into practical applications. Section 5 discusses the evaluation results, presenting a comparative analysis with the state-of-the-art ART4SQLi [9] across 10 test cases. Finally, Section 6 concludes the paper by summarizing the main findings and their implications.

## 2. Preliminaries

This section will introduce the fundamental concepts of software testing, test prioritization, and SQL injection, which are essential to understanding our proposed methodology.

### 2.1. Software Testing

Software testing encompasses any activity aimed at evaluating an attribute or capability of a program or system to determine whether it meets its requirements. Despite its critical role in ensuring software quality and its widespread use by programmers and testers, software testing remains largely an art due to the limited understanding of fundamental software principles. The inherent complexity of software makes comprehensive testing of even moderately complex programs infeasible. Testing extends beyond debugging; it serves various purposes, including quality assurance, verification and validation, and reliability assessment. Additionally, testing can function as a generic metric. Correctness testing and reliability testing represent two primary focus areas within software testing. Ultimately, software testing involves a trade-off between budget, time, and quality, necessitating carefully balancing these factors to achieve optimal results [10–12].

### 2.2. Test Prioritization

In order to reduce software testing costs, testers can prioritize test cases to ensure that the most critical ones are executed early in the testing process. This prioritization becomes particularly important during the maintenance and evolution phases, which are the stages where updates and changes are made to the software, of the software development life cycle, where regression testing is essential to confirm that recent changes have not adversely affected previous software versions, and that the new version remains backward compatible. Testing typically accounts for nearly half the total software development cost, making it a time-intensive and costly endeavor [10].

As sources [13–15] indicated, the regression testing process can be optimized through three primary methods: Test Case Selection, Test Suite Minimization, and Test Case Prioritization.

- Test Case Selection: This method selects test cases based on specific criteria, focusing only on the updated areas of the software.
- Test Suite Minimization: This approach reduces the overall test suite by eliminating redundant or obsolete test cases, optimizing testing resources, and minimizing the testing effort required.
- Test Case Prioritization: This technique involves ordering test cases based on specific attributes, ensuring that higher-priority test cases, which are those that cover critical functionalities or have a high likelihood of failure, are executed first [12,14,16,17].

Unlike Test Case Selection and Test Suite Minimization, which alter the original test suite by removing cases, Test Case Prioritization only reorders the test cases without eliminating any. This distinction is crucial because some test cases, though unnecessary for specific releases, may be valuable in future versions. As a result, prioritizing test cases instead of permanently removing them is often a safer strategy, providing a sense of security in the testing process. Therefore, Test Case Prioritization is considered a secure, reliable, and cost-effective approach to regression testing [18].

### 2.3. SQL Injection

SQL Injection is a prevalent network attack method [4,19]. It involves embedding malicious instructions within input strings to manipulate SQL syntax logic, aiming to unlawfully disrupt and infiltrate database servers. Such attacks can lead to the theft of confidential user information, account details, and passwords. We collected common SQL injection attack types from [19–24], detailed below:

- Boolean-based Blind SQL Injection: This method is one of the most common and dangerous types of SQL injection. It injects conditional SQL statements that force the SQL command to constantly evaluate as true, such as ('1' = '1'). This technique is primarily used to bypass authentication processes, allowing unauthorized access to the database.
- Union-based SQL Injection: In a union query attack, the attacker inserts an additional statement into the original SQL string. This injection is achieved by appending a UNION query string or a similar statement to a web form input. The result is that the database returns a dataset that combines the original query and the injected query results, thereby exposing additional data.
- Time-based Blind SQL Injection: This technique sends SQL queries that force the database to pause for a specified duration before responding. An attacker can infer information about the database by observing these delays, exploiting timing delays to gather data without direct feedback.
- Error-based SQL Injection: This attack depends on the error messages generated by the database server. These messages often reveal information about the database's structure, which can be used to craft further attacks.
- Stack-based SQL Injection: This involves injecting multiple SQL statements in a single query, separated by a delimiter such as a semicolon. The database executes these statements sequentially, allowing attackers to perform several actions with one injection.
- Inline Queries SQL Injection: Inline queries involve embedding malicious sub-queries within the main SQL query. These sub-queries can manipulate the database operation, often leading to unauthorized data access or manipulation.

## 3. Related Work

SQL injection (SQLi) remains a significant threat to the security of web applications, prompting the development of various detection and testing methodologies [21]. In 2019, Zhang et al. proposed ART4SQLi, an adaptive random testing method based on SQLMAP that prioritizes test cases to efficiently identify SQLi vulnerabilities, reducing the number of attempts needed by over 26% [9]. Unlike static order test prioritization, which calculates the distance between payloads, our tool employs a dynamic adjustment mechanism, enhancing detection efficiency.

Furthermore, most research needs to address the test prioritization of SQL injection. Al Wahaibi et al. introduced SQIRL in 2023, which utilizes deep reinforcement learning with grey-box feedback to intelligently fuzz input fields, generating diverse payloads, discovering more vulnerabilities with fewer attempts, and achieving zero false positives [22]. However, this method does not include test prioritization. Similarly, Erdődi et al. (2021) simulated SQLi attacks using Q-learning agents within a reinforcement learning framework, modeling SQLi as a security capture-the-flag challenge, enabling agents to learn generalizable attack strategies [25]. In the same year, Kasim developed an ensemble classification-based method that detects and classifies SQLi attacks as simple, unified, or lateral, utilizing features from the OWASP dataset to achieve high detection and classification accuracy [26]. Additionally, in 2023, Ravindran et al. created a Chrome extension to detect and prevent SQLi and XSS attacks by analyzing incoming data for suspicious patterns, thus enhancing web application security [2]. In 2024, Arasteh et al. presented a machine learning-based detection method using binary Gray-Wolf optimization for feature selection, which enhances detection efficiency by focusing on the most compelling features, achieving high accuracy and precision [23].

In test prioritization, Chen et al. investigated various techniques to optimize regression testing time in 2018. Based on test distribution analysis, their predictive test prioritization (PTP) method accurately predicts the optimal prioritization technique, significantly improving fault detection and reducing testing costs [27]. Haghighatkhah et al. studied the combination of diversity-based and history-based test prioritization (DBTP and HBTP) in continuous integration environments, finding that leveraging previous failure knowledge (HBTP) is highly effective. At the same time, DBTP is beneficial during early stages or when combined with HBTP [28]. Alptekin et al. introduced a method to prioritize security test executions based on web page similarities, hypothesizing that similar pages have similar vulnerabilities. This approach achieved high accuracy in predicting vulnerabilities, speeding up vulnerability assessments, and improving testing efficiency [29]. In 2023, Medeiros et al. proposed a clustering-based approach to categorize and prioritize code units based on security trustworthiness models. This method helps developers improve code security early in development by identifying code units prone to vulnerabilities, reducing potential vulnerabilities and associated costs [30].

These studies collectively advance the fields of SQL injection detection and test prioritization, providing robust methodologies to enhance web application security and optimize testing processes. However, to our knowledge, only ART4SQLi uses test prioritization to boost SQL injection testing.

## 4. Methodology and Implementation

The purpose of this section is to outline the methodology and steps undertaken in our research on testing prioritization for SQL injection vulnerabilities. The first section provides a succinct introduction to our penetration testing process and the associated framework. Subsequently, we detail the test prioritization algorithm in the following sections. Finally, the subsection introduces the profiling defense update function, elucidating its role in enhancing security measures.

### 4.1. TPSQLi Framework of Penetration Testing

Before initiating the detection module, various submodules targeting specific vulnerabilities are loaded. Figure 4 shows the workflow of our model.

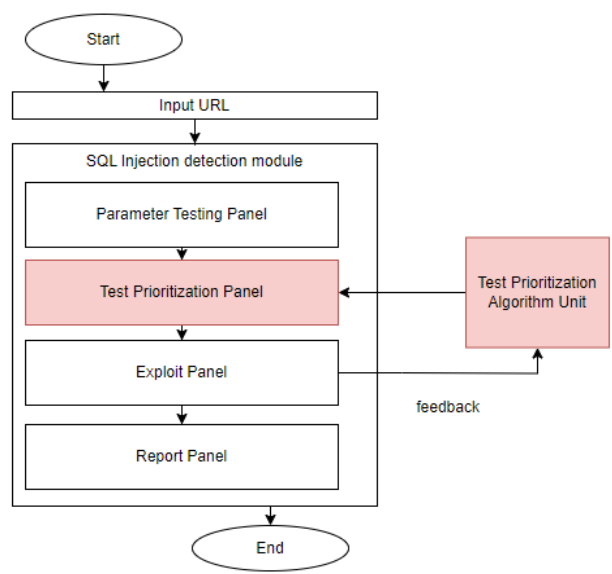

**Figure 4.** SQL injection attack detection model.

In this penetration framework, the initial step involves the user entering the URL to be tested. Subsequently, a web crawler is employed to identify additional test targets, with the crawling depth determined by user-defined settings. The next phase involves launching the SQL Injection Attack Detection Model, which encompasses four main components: the Parameter Testing Panel, Test Prioritization Panel, Exploit Panel, and Report Panel. In the Parameter Testing Panel, parameters can be transmitted using the GET or POST method. For GET requests, parameters are extracted directly from the web pages. HTML code is analyzed for POST requests to locate "form" tags. The subsequent Test Prioritization Panel calculates an appropriate testing sequence for the Exploit Panel, utilizing a test prioritization algorithm detailed in the following chapter. Once the parameters are extracted, they are tested based on the prioritization determined in the previous panel. The test results are then fed into the test prioritization algorithm unit to refine future testing sequences. Finally, the Report Panel displays the results of the detection process, provides relevant information, and offers recommended solutions.

*4.2. Parameter Testing Panel*

The Parameter Testing Panel involves extracting parameters for analysis using GET and POST, as referenced in [31]. The GET method extracts parameters following the "?" in the URL, while the POST method identifies parameters within form input tags for testing. This differentiation allows for a comprehensive analysis of all possible entry points for SQL injection vulnerabilities.

*4.3. Test Prioritization Panel*

Upon acquiring the test parameters, the Test Prioritization Panel employs a prioritization algorithm to determine the sequence of tests. The unit of the Test Prioritization Algorithm is shown in Figure 5. The calculated prioritization dictates the order in which tests are conducted. During the injection phase, the system dynamically adjusts the test prioritization using the profiling defense update function. Post-injection, feedback in feedback.json file is utilized to update the weight scores, ensuring the algorithm adapts to the testing environment. This iterative process refines the prioritization for subsequent tests, continuously optimizing the testing strategy.

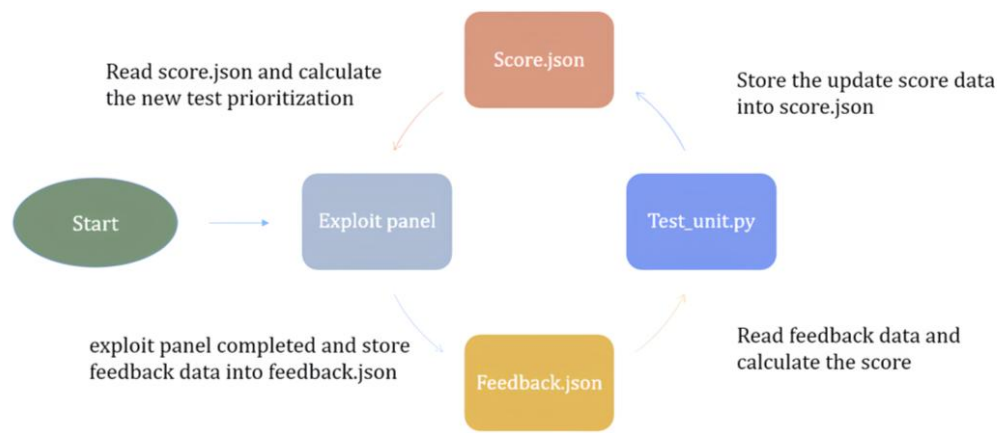

**Figure 5.** Our test prioritization algorithm unit.

*4.4. Test Prioritization Algorithm*

This subsection presents the test prioritization algorithm for SQL injection vulnerability detection in web applications.

4.4.1. Pseudocode

Algorithm 1 outlines the test prioritization algorithm for SQL injection vulnerability detection in web applications. The algorithm takes as input an $n$-dimensional Strength–Weakness (SW) vector, where $n$ represents different techniques, and a payload for each technique.

The process of our algorithm is detailed below:

1. Initialization (Set Initial Variables):
    a. Strength–Weakness Vector ('SW_Vector'): Initialize a fundamental score for each technique using a function discussed in the subsequent subsection. This vector quantifies each SQL injection technique's relative strengths or weaknesses based on historical data or predefined metrics.
    b. Exploit Time ('exploit_time'): Initialize this variable to record the fastest exploit time for each technique. For instance, if technique A exploits a vulnerability in 2 s, while technique A takes 5 s, this would influence the prioritization of these techniques.
    c. Is Exploit ('is_exploit'): Initialize this boolean variable to record the success status of each technique. If a technique fails, it will be marked as False, affecting its priority in future tests.

2. Execution Loop:
    a. Payload Selection: For each payload that failed in the previous round (higher risk payloads), determine whether it can be successfully exploited. If not, reduce its risk and select the payload with the highest fundamental score for testing.
    b. Execution: Start the execution loop by recording the start time to establish a baseline duration. Execute the selected payload, performing the specific task assigned by the system. After the payload execution, record the end time to calculate the total execution time, which is critical for performance analysis and optimization.
    c. Outcome Evaluation: Determine whether the exploit was successful. There are two possible outcomes:
    Failure Case:
        i. If 'is_exploit' for the technique is False, add the execution time and subtract one point from the fundamental score.
        ii. If 'is_exploit' is True, only subtract one point from the fundamental score without adding the execution time.
    Success Case:
        i. Set the payload's risk to high.
        ii. If 'is_exploit' for the technique is False, add the execution time and set 'is_exploit' to True.

3. Iteration and Update:
    a. Iterate: Continue testing the next payload until all payloads have been executed.
    b. Update: After executing all payloads, update the SW-vector based on the recorded exploit time and the 'is_exploit' status.

Our proposed algorithm ensures that the payloads with higher risks are prioritized and the execution times are minimized by continuously updating and adapting the strength–weakness vector based on the performance of each technique. For an example of the algorithm in action, please refer to Section 4.4.4.

**Algorithm 1** Test Prioritization Algorithm

**Input:** Strength–Weakness Vector SW_vector$[0, \ldots, n-1]$, Payloads payloads $= \{0 : [\ldots], 1 : [\ldots], \ldots, n-1 : [\ldots]\}$

1: Initialize Basic Scores BS[0, …, n-1] based on SW_vector
2: Initialize exploit_time[0, …, n-1] to zero
3: Initialize is_exploit[0, …, n-1] to False
4: **while** there exists an unused payload with risk level 3 **do**
5:    Select payload $p$ with the maximum BS[t] where $\text{risk}(p) = 3$ and $t \in [0, \ldots, n-1]$
6:    **if** $p$ fails to exploit **then**
7:       Set $\text{risk}(p) \leftarrow 1$
8:    **end if**
9: **end while**
10: **while** there are unused payloads remaining **do**
11:    Select payload $p$ with the maximum BS[t] where $t \in [0, \ldots, n-1]$
12:    **Record** current time as start_time
13:    **Execute** payload $p$
14:    **Record** current time as end_time
15:    **if** $p$ fails to exploit **then**
16:       **if** is_exploit$[t] = $ False **then**
17:          exploit_time$[t] \leftarrow$ exploit_time$[t] + ($end_time $-$ start_time$)$
18:          BS[t] $\leftarrow$ BS[t] $- 1$
19:       **else**
20:          Set $\text{risk}(p) \leftarrow 3$
21:          **if** is_exploit$[t] = $ False **then**
22:             exploit_time$[t] \leftarrow$ exploit_time$[t] + ($end_time $-$ start_time$)$
23:             Set is_exploit$[t] \leftarrow$ True
24:          **end if**
25:       **end if**
26:    **end if**
27: **end while**
28: **Update** SW_vector based on exploit_time[0, …, n-1] and is_exploit[0, …, n-1]

4.4.2. Mathematical Formulation of Test Prioritization Algorithm

In this subsection, we introduce the algorithm for test prioritization of various SQL injection vulnerability detection techniques. We define two sets: one representing all current detection techniques and the other representing the set of test targets. As technology evolves, new techniques may emerge. We assume there are currently $n$ detection techniques and $m$ test targets for this discussion.

For each test target, we conduct $n$ attacks and record the duration of these attacks, resulting in $n \times m$ different results, which are then used to calculate the weight scores. We establish three rules to construct this test prioritization model:

- Rule 1: If a technique fails to exploit a vulnerability successfully, it receives zero points in the weight score calculation.
- Rule 2: The total score for the same test target is $n$, with faster exploits earning higher scores.
- Rule 3: The score calculation considers the difference in exploit time using the reciprocal ratio of these times.

Moreover, we use the following mathematical formulas to express the model:

- Techniques: $n$.
- Successful exploits: $a$.
- Failed exploits: $n - a$.
- Success exploit time: $S_1, S_2, \ldots, S_a$.
- Fail exploit time: $F_1, F_2, \ldots, F_a$.

- Weights (Both $W_{F_i}, W_{S_i}, W_i$ set to zero in the initial process):

$$W_{F_i} = 0, \forall i = 1,2,\ldots, n - a \quad (1)$$

$$W_{S_i} = \frac{1}{S_i} \cdot n \cdot \frac{1}{\sum_{k=1}^{a} \frac{1}{S_k}}, \forall i = 1,2,\ldots, a \quad (2)$$

$$W_i = \max(W_{S_i}, W_{F_i}), \forall i = 1,2,\ldots, a \quad (3)$$

For a deeper understanding of our algorithm, please refer to Section 4.4.4 for a simple example.

4.4.3. Profiling Defense-Update Function

In addition to adjusting the order of the basic technique through the previously mentioned algorithm, we have incorporated an update mechanism within the dynamic segment. This mechanism aims to facilitate the dynamic adjustment of test prioritization. Specifically, after each technique is employed with an injected payload, if no vulnerabilities are detected, the priority of this technique will be progressively reduced in subsequent tests. Concurrently, we also prioritize the implementation of high-weighted techniques.

The precise methodology involves leveraging the fundamental score calculated by the preceding algorithm. When a test is conducted, and no weaknesses are identified, the technique score is decremented. This decrement continues until the score falls below that of other techniques or until the technique is deemed ineffective. This adaptive scoring and prioritization process ensures that the focus gradually shifts towards techniques more likely to uncover vulnerabilities, thereby optimizing the efficiency and effectiveness of the testing procedure.

4.4.4. Simple Example for Test Prioritization Algorithm

To better understand our algorithm, consider a simplified scenario involving three SQL injection techniques: Technique A1, Technique A2, and Technique A3. If, in the first round, Technique A1 has an exploit time of 2 s, Technique A2 fails, and Technique A3 has an exploit time of 4 s, the weight calculations would be as follows:
- Technique A1: $W_i = 2$
- Technique A2: $W_i = 0$
- Technique A3: $W_i = 1$

Given these weights, Technique A1 has the highest priority. Therefore, our method will adjust the profiling defense-update function to prioritize testing Technique A1 first, followed by Technique A3, and finally Technique A2.

*4.5. Exploit Panel*

As shown in Figure 6, the Exploit Panel initiates by calculating a fundamental score using a strength–weakness vector. Two variables are initialized: 'exploit_time' to record the fastest exploit time for each technique and 'is_exploit' to track the success of each technique. The panel first re-tests previously failed payloads, adjusting their risk levels dynamically based on success or failure. Successful exploits prompt the setting of the payload's risk to high and update the technique's status.

For failed payloads, the system subtracts points from their fundamental score and re-tests the next payload. If a payload is successfully exploited, its execution time is recorded, and the strength–weakness vector is updated accordingly. This process continues until all payloads are tested, ensuring comprehensive coverage of potential vulnerabilities.

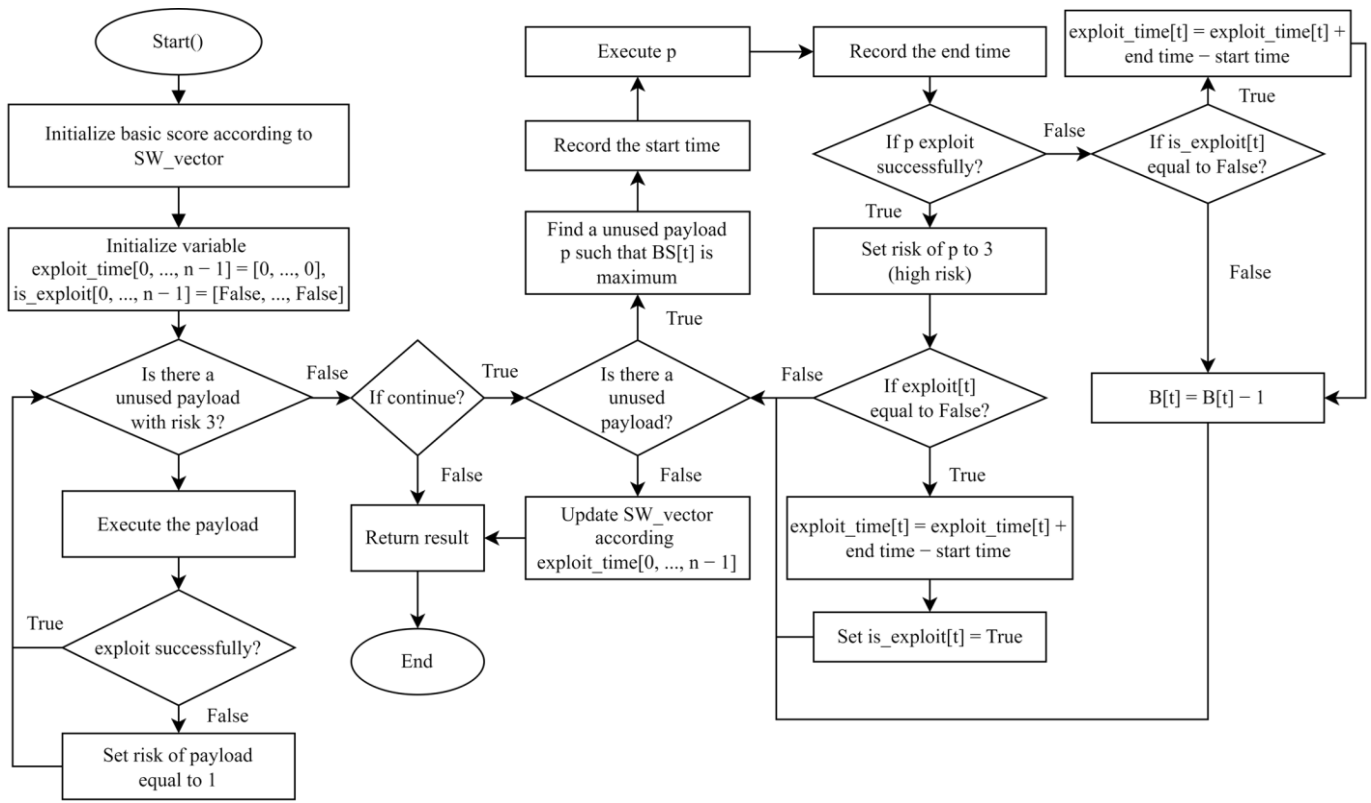

**Figure 6** Flowchart of Exploit panel.

*4.6. Report Panel*

　　Upon completing all tests, the system generates a comprehensive report, concluding the penetration testing process. The described implementation ensures a systematic and thorough approach to detecting SQL injection vulnerabilities in web applications.

## 5. Evaluation

　　The experimental setup included a Windows 10 machine from ASUS, model P2520LA, which was produced in China. The machine equipped with 12 GB of RAM and an Intel(R) Core(TM) i5-5200U CPU, operating at 2.2 GHz with four cores. The penetration testing framework, as outlined in Section 4, was implemented using Python 3.9.7. Our study focused on two specific targets, DVWA SQL-blind and DVWA SQL, along with eight real-world cases, including login pages, blogs, business websites, and eCommerce sites. Our framework was first evaluated on an open-source software project, with the specific test targets depicted in Table 2.

**Table 2.** TPSQLi test target.

| Type | Test Target | Topic |
|---|---|---|
| Open Source | DVWA (SQL-blind) | Test Environment |
|  | DVWA (SQL) | Test Environment |
| Real-world Case | R1 | Login page |
|  | R2 | Wiki/database |
|  | R3 | Blogs/news |
|  | R4 | Business website |
|  | R5 | Service provider |
|  | R6 | eCommerce website |
|  | R7 | Portfolio |
|  | R8 | Business website |

## 5.1. Collecting Time Data from Various Techniques

This flowchart in Figure 7 is designed to collect time data for various SQL injection detection techniques. The process begins by inputting a URL as the test target and testing each technique over multiple rounds ($n$ rounds). The technique type is recorded at the outset, followed by the start time for each technique test. The exploit is then executed, and upon completion, the end time is recorded regardless of the test's success. The log file is subsequently deleted to ensure it does not affect subsequent tests. This procedure is repeated for all techniques, ensuring a comprehensive time data collection for each method.

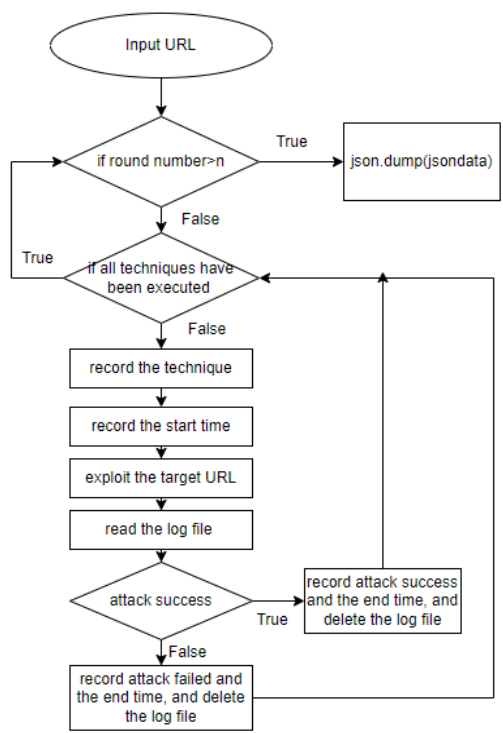

**Figure 7.** Flowchart of collecting time data using various techniques.

## 5.2. Weights for Test

We evaluated TPSQLi on two open-source projects and eight websites with publicly disclosed SQL injection vulnerabilities submitted to HITCON ZeroDay [32]. The types and themes of these test websites are detailed in Table 2.

**Table 3.** Weights of the test target.

|  | Boolean-Based | Error-Based | Union-Based | Stack-Based | Time-Based | Inline Queries |
|---|---|---|---|---|---|---|
| DVWA (SQL-blind) | 5.46 | 0 | 0 | 0.25 | 0.29 | 0 |
| DVWA (SQL) | 2.34 | 0.94 | 2.57 | 0.09 | 0.06 | 0 |
| R1 | 0 | 0 | 0 | 0 | 6 | 0 |
| R2 | 2.32 | 0 | 2.55 | 0 | 1.13 | 0 |
| R3 | 0 | 0 | 0 | 0 | 6 | 0 |
| R4 | 6 | 0 | 0 | 0 | 0 | 0 |
| R5 | 4.59 | 0 | 0 | 0.43 | 0.98 | 0 |
| R6 | 3.60 | 1.23 | 0.75 | 0 | 0.42 | 0 |
| R7 | 2.71 | 0 | 2.54 | 0 | 0.74 | 0 |
| R8 | 2.74 | 0.69 | 2.16 | 0 | 0.40 | 0 |

We performed five rounds of testing, recording the results and calculating the weights using Equations (1) and (2). The compiled scores are presented in Table 3. These scores informed the development of new testing priorities, transitioning from the existing BEUSTQ to a new order based on the calculated weights. Detailed timing information for each round is available in Appendix A.

*5.3. Comparing with ART4SQLi*

To evaluate the effectiveness of our proposed test prioritization approach, we conducted a comparative analysis against ART4SQLi, a widely used tool for SQL injection detection.

5.3.1. Coverage-Based Visual Comparison

We plotted the data obtained from both the original test prioritization and our new test prioritization approach on a single chart for visual comparison. In each chart in Figure 8, the horizontal axis represents time, while the vertical axis denotes the number of detected vulnerabilities. The blue area illustrates the results from the original test order, whereas the yellow area represents the outcomes generated by our TPSQLi framework. Our research demonstrates that the test priorities determined by the TPSQLi framework consistently outperform or match the original test order. In scenarios where the test priority differs from the original order, our approach achieves faster and more remarkable coverage improvement, leading to earlier attainment of 100% coverage.

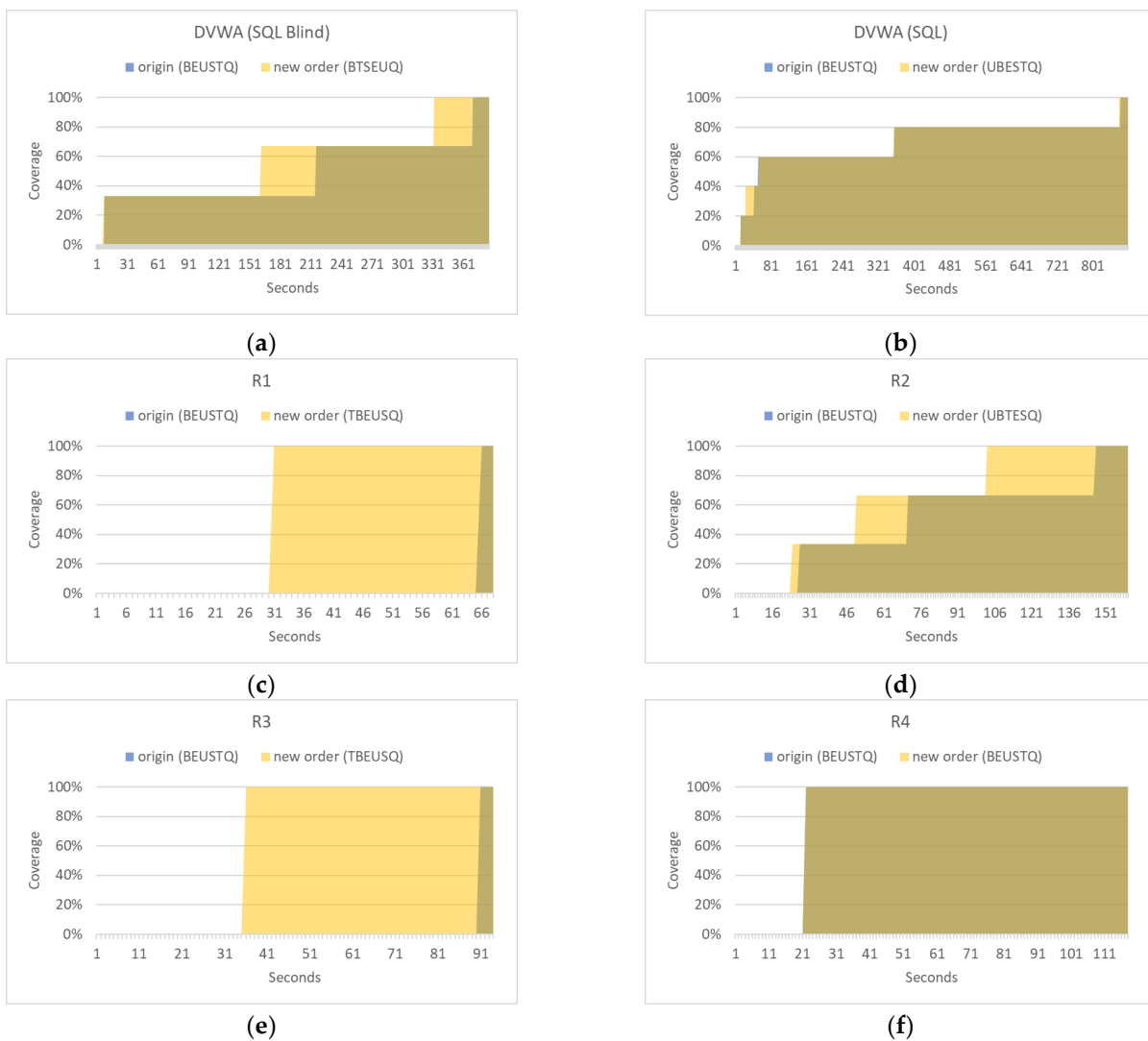

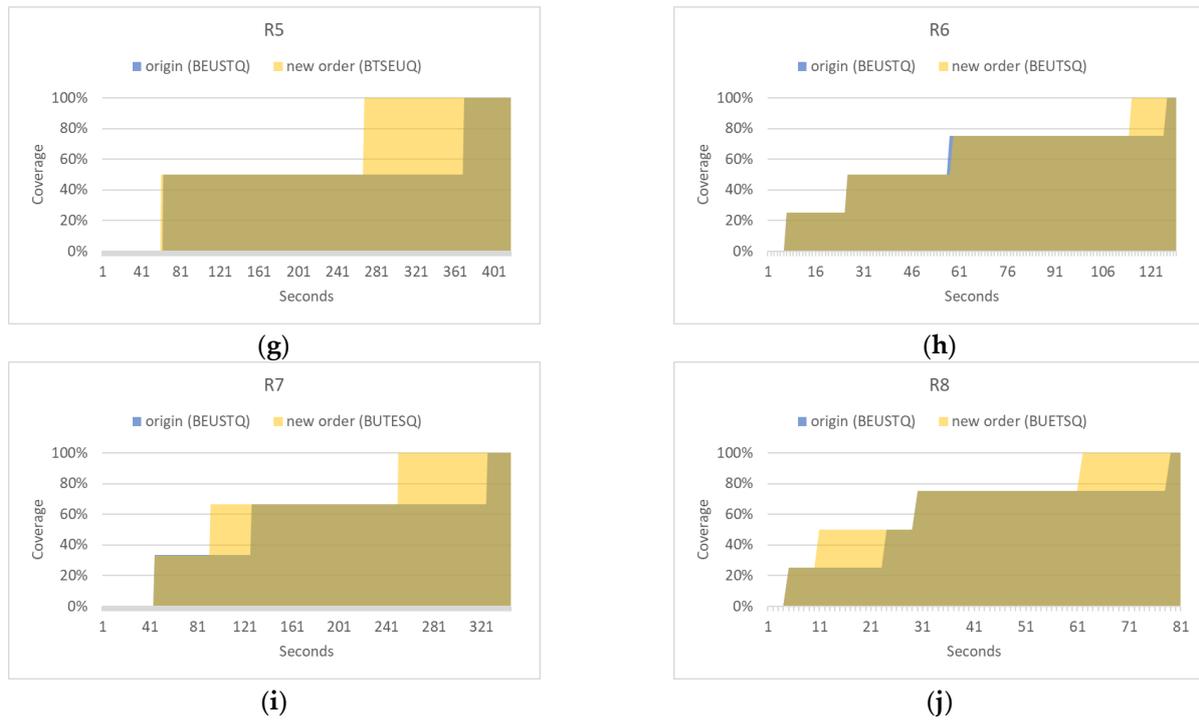

**Figure 8** Comparison between the original tool (ART4SQLi) and the optimized tool (TPSQLi) was conducted across various cases: (**a**) DVWA (SQL blind); (**b**) DVWA (SQL); (**c**) R1; (**d**) R2; (**e**) R3; (**f**) R4; (**g**) R5; (**h**) R6; (**i**) R7; (**j**) R8.

5.3.2. Statistical Analysis of Test Execution Time

The maximum time $T_{max}$, minimum time $T_{min}$, average time $T_{mean}$, and standard deviation of the time for exposing the last vulnerabilities $T_{std}$ of both the ART4SQLi and the proposed TPSQLi for the ten test targets are presented in Table 4.

**Table 4.** Comparisons between ART4SQLi and TPSQLi.

| Test Target | Method | $T_{max}$ | $T_{min}$ | $T_{mean}$ | $T_{std}$ | $|Z|$ |
|---|---|---|---|---|---|---|
| DVWA (SQL-blind) | ART4SQLi [9] | 405.80 | 349.90 | 369.11 | 20.37 | 1.68 |
|  | TPSQLi | **359.9** | **307.06** | **331.26** | **14.47** |  |
| DVWA (SQL) | ART4SQLi [9] | 913.55 | 834.79 | 862.39 | 31.95 | 0.059 |
|  | TPSQLi | **898.65** | **836.08** | **859.97** | **25.52** |  |
| R1 | ART4SQLi [9] | 68.48 | 64.26 | 65.82 | 1.69 | 17.04 |
|  | TPSQLi | **32.42** | **29.28** | **30.44** | **1.21** |  |
| R2 | ART4SQLi [9] | 157.61 | 141.22 | 146.07 | 6.76 | 5.50 |
|  | TPSQLi | **109.82** | **99.45** | **102.73** | **4.07** |  |
| R3 | ART4SQLi [9] | 94.11 | 87.41 | 90.68 | 2.65 | 20.66 |
|  | TPSQLi | **35.69** | **34.67** | **35.15** | **0.46** |  |
| R4 | ART4SQLi [9] | 22.63 | 20.10 | **21.00** | 0.99 | 0.04 |
|  | TPSQLi | **22.47** | **20.32** | 21.06 | **0.85** |  |
| R5 | ART4SQLi [9] | 395.77 | 351.55 | 368.99 | 16.29 | 5.86 |
|  | TPSQLi | **275.34** | **259.71** | **267.65** | **5.83** |  |
| R6 | ART4SQLi [9] | 128.62 | 121.16 | 124.49 | 3.10 | 2.67 |
|  | TPSQLi | **116.81** | **109.65** | **113.21** | **2.88** |  |
| R7 | ART4SQLi [9] | 332.98 | 319.94 | 325.47 | 5.69 | 9.49 |
|  | TPSQLi | **256.94** | **244.99** | **250.87** | **5.43** |  |
| R8 | ART4SQLi [9] | 79.27 | 78.18 | 78.72 | 0.44 | 32.16 |
|  | TPSQLi | **61.90** | **61.15** | **61.63** | **0.30** |  |

The last column in Table 4, labeled $|Z|$, displays the values obtained from the statistical *Z-test*, calculated using the mean ($T_{mean}$) and standard deviation ($T_{std}$), specified in Equation (4). This column is intended to determine whether there is a statistically significant difference between the results of the two approaches (ART4SQLi and TPSQLi) across the ten test cases, with a confidence level of 95%, i.e., $Z_{0.95} = 1.645$.

$$Z = \frac{T_{mean,TPSQLi} - T_{mean,ART4SQLi}}{\sqrt{T_{std,TPSQLi}^2 + T_{std,ART4SQLi}^2}} \quad (4)$$

As shown in Table 4, TPSQLi outperformed ART4SQLi across all metrics, with the exception of $T_{mean}$ for R4. Notably, the test prioritization did not alter the order for R4 and DVWA (SQL), resulting in the same order as observed in ART4SQLi. The $|Z|$ values, except for those corresponding to DVWA (SQL) and R4, exceeded the critical value $Z_{0.95} = 1.645$, as indicated in the rightmost column. This suggests that the *Z-test* confirms a significant reduction in execution time between ART4SQLi and TPSQLi for the 10 test cases.

5.3.3. Discussion of Effectiveness

Like ART4SQLi [9], we discuss the effectiveness of our test prioritization process and compare it with ART4SQLi. In TPSQLi, when a payload successfully exposes a vulnerability in the web application, it is considered valid and treated as a true positive. If the payload detects the final vulnerability, TPSQLi halts, marking the payload as valid. All payloads tested before this last successful one, which did not lead to exploiting a vulnerability, are considered false positives (they are flagged as suspicious but ultimately invalid).

If TPSQLi stops after the $p^{th}$ payload (where $p \geq 1$ and the payload is valid), and the number of the valid payloads before $p^{th}$ is $v$, the $p - v - 1$ payloads are categorized as false positives. In contrast, the $v + 1$ payloads are considered as true positive. Since TPSQLi terminates once the last valid payload is detected, neither true nor false negatives are accounted for in this analysis.

We designed a new metric, the False Positive Measure (*FPM*), to systematically compare the false positives. This metric is defined as:

$$FPM = \frac{Total\ Executed\ Payloads}{False\ Positive} \quad (5)$$

A higher $FPM$ indicates a more effective process, implying fewer false positives than the number of executed payloads. To further quantify the improvement of TPSQLi over ART4SQLi, we also introduce an Improved Rate (*IR*), calculated as:

$$IR = \frac{FPM_{TPSQLi} - FPM_{ART4SQLi}}{FPM_{ART4SQLi}} \times 100\% \quad (6)$$

The *IR* expresses the percentage improvement offered by TPSQLi over ART4SQLi. As shown in Table 5, TPSQLi delivers significant enhancements, particularly for test targets R1 and R3, with *IR* values of 19.95% and 16.62%, respectively. On average, TPSQLi demonstrates a 4.65% overall improvement, indicating consistent efficiency gains across different web applications. Notably, no improvement was observed on test targets DVWA(SQL) and R4, as the original order in both cases was optimal, resulting in identical *FPM* values for both TPSQLi and ART4SQLi.

This analysis highlights that our test prioritization method effectively reduces false positives and improves the accuracy of SQL injection detection. The increase in *FPM* values and positive IR percentages confirm that TPSQLi is superior to ART4SQLi, particularly in handling test execution time and precision.

**Table 5.** Comparison of TPSQLi and ART4SQLi Based on False Positive Measure (*FPM*) and Improved Rate (*IR*) Across Various Test Targets.

| Test Target | Method | *FPM* | *IR* | Average *IR* |
|---|---|---|---|---|
| DVWA (SQL-blind) | ART4SQLi [9] | 1.0028 | 3.37% | 4.65% |
| | TPSQLi | 1.0366 | | |
| DVWA (SQL) | ART4SQLi [9] | 1.0057 | 0.00% | |
| | TPSQLi | 1.0057 | | |
| R1 | ART4SQLi [9] | 1.0004 | 19.95% | |
| | TPSQLi | 1.2000 | | |
| R2 | ART4SQLi [9] | 1.0018 | 2.17% | |
| | TPSQLi | 1.0236 | | |
| R3 | ART4SQLi [9] | 1.0004 | 16.62% | |
| | TPSQLi | 1.1667 | | |
| R4 | ART4SQLi [9] | 1.0004 | 0.00% | |
| | TPSQLi | 1.0004 | | |
| R5 | ART4SQLi [9] | 1.0009 | 0.71% | |
| | TPSQLi | 1.0080 | | |
| R6 | ART4SQLi [9] | 1.0031 | 1.09% | |
| | TPSQLi | 1.0140 | | |
| R7 | ART4SQLi [9] | 1.0018 | 1.70% | |
| | TPSQLi | 1.0189 | | |
| R8 | ART4SQLi [9] | 1.0030 | 0.92% | |
| | TPSQLi | 1.0123 | | |

## 6. Conclusions

In this research, we propose a comprehensive framework for regression testing of SQLi, focused on optimizing test prioritization through the design of weight functions and evaluation methods. The proposed module for calculating test prioritization is flexible, accommodating various technologies and targets, and includes dynamic adjustment capabilities. The weight calculation method accounts for the time differential in the exposure of weaknesses and considers the probability of successful exploitation by different technologies. Our findings demonstrate that the time to expose the first weakness is significantly reduced, and the overall exposure time is faster than the original test order. TPSQLi effectively accelerates the testing process, as evidenced by statistical *Z-test* calculations confirming significant differences compared to the ART4SQLi. We also discuss the false positive to check the effectiveness of TPSQLi and compare it with ART4SQLi. Moreover, future work will explore machine learning methods, such as large language models for feature extraction and reinforcement learning for dynamically adjusting test prioritization, to enhance test prioritization further, aiming to improve the efficiency of SQL Injection testing.

**Author Contributions:** Conceptualization, F.W., G.-Y.Y. and Y.-Z.G.; data curation, G.-Y.Y.; formal analysis, G.-Y.Y.; investigation, Y.-Z.G., G.-Y.Y. and F.W.; methodology, G.-Y.Y. and Y.-Z.G.; implementation and experiment, Y.-Z.G., G.-Y.Y., and P.-H.H.; visualization, G.-Y.Y.; writing—original draft preparation, G.-Y.Y.; writing—review and editing, G.-Y.Y., Y.-W.T., K.-H.Y., F.W. and W.-L.W., supervision, F.W., G.-Y.Y. and K.-H.Y.; project administration, F.W. and K.-H.Y. All authors have read and agreed to the published version of the manuscript.

**Funding:** This research was partly supported by the National Science and Technology Council (NSTC), Taiwan, ROC, under the projects MOST 110-2221-E-002-069-MY3, NSTC 111-2221-E-A49-202-MY3, NSTC 112-2634-F-011-002-MBK and NSTC 113-2634-F-011-002-MBK.. We also received partial support from the 2024 CITI Visible Project: Questionnaire-based Technology for App Layout Evaluation, Academia Sinica, Taiwan, ROC.

**Institutional Review Board Statement:** Not applicable.

**Informed Consent Statement:** Not applicable.

**Data Availability Statement:** The data presented in the study are included in the article, further inquiries can be directed to the corresponding author/s.

**Acknowledgments:** We thank the anonymous reviewers for their valuable comments, which make this manuscript better. We also thank Jui-Ning Chen from Academia Sinica, Taiwan, for her invaluable comments. We appreciate the Speech AI Research Center of National Yang Ming Chiao Tung University for providing the necessary computational resources. Additionally, we utilized GPT-4 to assist with wording, formatting, and stylistic improvements throughout this research.

**Conflicts of Interest:** The authors declare no conflict of interest.

## Appendix A

The unit of time data is seconds.

**Table A1.** Time data of various techniques (DVWA SQL-blind).

|         | Boolean-Based | Error-Based | Union-Based | Stack-Based | Time-Based | Inline Queries |
|---------|---------------|-------------|-------------|-------------|------------|----------------|
| 1       | 4.42          | 23.08       | 22.09       | 119.07      | 195.62     | 21.34          |
| 2       | 4.23          | 4.01        | 22.11       | 253.69      | 296.76     | 21.39          |
| 3       | 4.19          | 3.90        | 22.14       | 215.18      | 61.29      | 21.29          |
| 4       | 4.94          | 23.07       | 22.97       | 100.12      | 118.80     | 22.13          |
| 5       | 23.34         | 23.00       | 3.07        | 195.92      | 99.64      | 2.186          |
| Average | 8.23          | 15.40       | 14.26       | 176.80      | 154.42     | 17.67          |
| Score   | 5.46          | 0           | 0           | 0.25        | 0.29       | 0              |

**Table A2.** Time data of various techniques (DVWA SQL).

|         | Boolean-Based | Error-Based | Union-Based | Stack-Based | Time-Based | Inline Queries |
|---------|---------------|-------------|-------------|-------------|------------|----------------|
| 1       | 8.44          | 43.41       | 3.34        | 272.56      | 541.80     | 21.60          |
| 2       | 8.30          | 21.59       | 22.51       | 215.04      | 464.64     | 2.36           |
| 3       | 8.55          | 22.05       | 3.46        | 196.03      | 783.58     | 21.57          |
| 4       | 26.90         | 41.05       | 3.47        | 311.27      | 387.10     | 21.54          |
| 5       | 8.47          | 23.24       | 22.57       | 522.80      | 349.79     | 21.56          |
| Average | 12.13         | 30.27       | 11.07       | 303.54      | 505.38     | 17.73          |
| Score   | 8.44          | 43.41       | 3.34        | 272.56      | 541.80     | 21.60          |

**Table A3.** Time data of various techniques (R1).

|         | Boolean-Based | Error-Based | Union-Based | Stack-Based | Time-Based | Inline Queries |
|---------|---------------|-------------|-------------|-------------|------------|----------------|
| 1       | 3.84          | 4.13        | 22.43       | 4.85        | 29.78      | 2.25           |
| 2       | 3.88          | 4.18        | 21.67       | 5.16        | 30.00      | 2.14           |
| 3       | 3.94          | 4.34        | 21.14       | 5.15        | 29.69      | 2.21           |
| 4       | 4.04          | 4.54        | 21.25       | 5.98        | 30.63      | 2.19           |
| 5       | 4.07          | 4.28        | 21.92       | 5.79        | 32.42      | 2.15           |
| Average | 3.95          | 4.29        | 21.68       | 5.39        | 30.51      | 2.19           |
| Score   | 0             | 0           | 0           | 0           | 6          | 0              |

**Table A4.** Time data of various techniques (R2).

|   | Boolean-Based | Error-Based | Union-Based | Stack-Based | Time-Based | Inline Queries |
|---|---|---|---|---|---|---|
| 1 | 25.11 | 19.53 | 23.82 | 20.85 | 52.53 | 13.94 |
| 2 | 25.97 | 20.22 | 22.04 | 21.91 | 53.11 | 13.2 |
| 3 | 24.4 | 20.13 | 22.76 | 21.19 | 52.74 | 14.06 |
| 4 | 26.21 | 20.88 | 22.49 | 23.81 | 53.02 | 14.35 |
| 5 | 28.84 | 21.38 | 27.46 | 24.41 | 55.52 | 13.90 |
| Average | 26.10 | 20.43 | 23.71 | 22.44 | 53.39 | 13.90 |
| Score | 2.32 | 0 | 2.55 | 0 | 1.13 | 0 |

**Table A5.** Time data of various techniques (R3).

|   | Boolean-Based | Error-Based | Union-Based | Stack-Based | Time-Based | Inline Queries |
|---|---|---|---|---|---|---|
| 1 | 7.71 | 8.48 | 27.43 | 9.07 | 34.72 | 4.05 |
| 2 | 7.62 | 8.89 | 28.92 | 9.11 | 34.78 | 4.42 |
| 3 | 7.74 | 9.33 | 29.87 | 9.9 | 35.69 | 3.93 |
| 4 | 7.23 | 10.11 | 31.63 | 8.6 | 36.54 | 4.16 |
| 5 | 7.46 | 8.45 | 30.02 | 9.06 | 35.09 | 4.04 |
| Average | 7.55 | 9.05 | 29.57 | 9.15 | 35.36 | 4.12 |
| Score | 0 | 0 | 0 | 0 | 6 | 0 |

**Table A6.** Time data of various techniques (R4).

|   | Boolean-Based | Error-Based | Union-Based | Stack-Based | Time-Based | Inline Queries |
|---|---|---|---|---|---|---|
| 1 | 22.63 | 17.21 | 27.44 | 19.74 | 20.96 | 11.72 |
| 2 | 20.32 | 17.7 | 26.74 | 18.79 | 20.54 | 12.04 |
| 3 | 21.09 | 17.88 | 28.78 | 19.91 | 21.2 | 12.19 |
| 4 | 20.1 | 16.59 | 27.06 | 19.08 | 21.29 | 10.91 |
| 5 | 20.89 | 16.26 | 27.08 | 19.81 | 20.8 | 12.21 |
| Average | 21.00 | 17.13 | 27.42 | 19.47 | 20.96 | 11.81 |
| Score | 6 | 0 | 0 | 0 | 0 | 0 |

**Table A7.** Time data of various techniques (R5).

|   | Boolean-Based | Error-Based | Union-Based | Stack-Based | Time-Based | Inline Queries |
|---|---|---|---|---|---|---|
| 1 | 65.59 | 17.59 | 78.19 | 202.49 | 42.81 | 5.22 |
| 2 | 61.43 | 20.44 | 81.52 | 201.87 | 44.99 | 5.56 |
| 3 | 63.91 | 17.72 | 87.58 | 199.29 | 35.91 | 5.61 |
| 4 | 58.72 | 17.54 | 77.35 | 197.94 | 38.85 | 5.78 |
| 5 | 62.78 | 27.95 | 92.36 | 212.68 | 46.66 | 12.93 |
| Average | 62.49 | 20.25 | 83.40 | 202.85 | 41.85 | 7.02 |
| Score | 4.59 | 0 | 0 | 1.41 | 0 | 0 |

**Table A8.** Time data of various techniques (R6).

|   | Boolean-Based | Error-Based | Union-Based | Stack-Based | Time-Based | Inline Queries |
|---|---|---|---|---|---|---|
| 1 | 6.63 | 18.05 | 31.0 | 10.68 | 55.29 | 3.05 |
| 2 | 6.41 | 20.02 | 32.07 | 12.26 | 57.86 | 3.04 |
| 3 | 6.33 | 18.0 | 31.15 | 10.46 | 55.22 | 3.4 |
| 4 | 6.86 | 19.16 | 30.96 | 11.23 | 57.39 | 3.18 |
| 5 | 6.47 | 20.09 | 31.83 | 11.22 | 55.86 | 3.16 |
| Average | 6.54 | 19.06 | 31.40 | 11.17 | 56.32 | 3.17 |
| Score | 3.60 | 1.23 | 0.75 | 0 | 0.42 | 0 |

**Table A9.** Time data of various techniques (R7).

| | Boolean-Based | Error-Based | Union-Based | Stack-Based | Time-Based | Inline Queries |
|---|---|---|---|---|---|---|
| 1 | 44.58 | 34.07 | 47.84 | 36.67 | 157.65 | 21.53 |
| 2 | 43.0 | 36.19 | 45.25 | 37.28 | 158.22 | 21.55 |
| 3 | 42.37 | 33.57 | 48.07 | 38.94 | 166.8 | 21.48 |
| 4 | 45.07 | 35.82 | 48.15 | 41.16 | 162.78 | 22.65 |
| 5 | 45.26 | 35.55 | 45.49 | 38.41 | 159.21 | 15.5 |
| Average | 44.05 | 35.04 | 46.96 | 38.49 | 160.93 | 20.54 |
| Score | 2.71 | 0 | 2.54 | 0 | 0.74 | 0 |

**Table A10.** Time data of various techniques (R8).

| | Boolean-Based | Error-Based | Union-Based | Stack-Based | Time-Based | Inline Queries |
|---|---|---|---|---|---|---|
| 1 | 4.92 | 18.3 | 5.95 | 17.55 | 32.11 | 2.62 |
| 2 | 4.69 | 19.09 | 5.98 | 17.49 | 31.7 | 2.73 |
| 3 | 4.54 | 18.38 | 6.24 | 17.12 | 32.1 | 2.32 |
| 4 | 4.67 | 18.53 | 6.01 | 17.19 | 31.78 | 2.37 |
| 5 | 4.71 | 19.16 | 5.66 | 17.65 | 32.09 | 2.43 |
| Average | 4.70 | 18.69 | 5.97 | 17.40 | 31.96 | 2.49 |
| Score | 2.74 | 0.69 | 2.16 | 0 | 0.40 | 0 |